\documentclass[a4paper,11pt]{article}
\usepackage{pos}
\usepackage{aas_macros}
\usepackage{siunitx}
\usepackage[UKenglish]{babel}
\usepackage{subcaption}

\DeclareSIUnit\year{yr}
\DeclareSIUnit\parsec{pc}

\title{Diffuse emission from stochastic sources}

\author*[a]{Anton Stall}
\author[a]{Philipp Mertsch}

\affiliation[a]{Institute for Theoretical Particle Physics and Cosmology (TTK), RWTH Aachen University, \\
52056 Aachen, Germany}

\emailAdd{stall@physik.rwth-aachen.de}
\emailAdd{pmertsch@physik.rwth-aachen.de}

\abstract{Diffuse emission in gamma-rays and neutrinos are produced by the interaction of cosmic rays with the interstellar medium. Below some hundreds of TeV, the sources of these cosmic rays are most likely Galactic. Hence, observations of high-energy gamma-rays and neutrinos can be used to probe the flux of cosmic rays in other parts of the Galaxy. Supernova remnants are usually considered as the prime candidate for the acceleration of Galactic cosmic rays. They inject cosmic rays in a point-like and specific time-dependent manner. As the precise positions and ages of the sources are not known, predictions must be obtained in a stochastic model. At GeV energies, the distribution of sources can be approximated with a smoothly varying spatial and temporal source density. At hundreds of TeV, however, the point-like nature matters as less sources contribute effectively due to shorter escape times. We have modelled diffuse emissions at hundreds of TeV, relevant for measurements by LHAASO, Tibet AS-gamma, IceCube, and the upcoming SWGO, as well as at tens of GeV, as measured by Fermi-LAT. This reveals the distinctive nature of diffuse emissions at the respective energies which can likely be used to constrain source models and locate cosmic ray sources.}

\ConferenceLogo{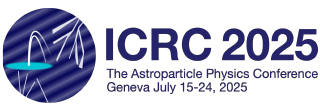}

\FullConference{39th International Cosmic Ray Conference (ICRC2025)\\
 15–24 July 2025\\
Geneva, Switzerland\\}

\begin{document}
\maketitle

\section{Introduction\label{sec:introduction}}
The discovery of cosmic rays (CRs) in the early 20th century raised fundamental questions that still persist:
What are their sources in our Galaxy?
Where are these sources?
And how do Galactic CRs reach energies as high as the so-called CR knee at around \SI{3}{\peta\electronvolt}, marking the transition from CRs of Galactic to extra-Galactic origin?
Thanks to experiments like AMS-02~\cite{2021AguilarAliCavasonzaPhR}, DAMPE~\cite{2024AlemannoAltomarePhRvD}, and the proposed AMS-100~\cite{2019SchaelAtanasyanNIMPA}, we now have access to increasingly precise measurements of the CR spectrum at Earth, potentially revealing spectral features hinting at the nature of CR sources (see, e.g.,~\cite{2025StallLooApJL}).
However, these observations are limited to our specific position in the Galaxy .

An alternative approach to study CRs is the analysis of Galactic diffuse emissions (GDEs) generated by CR interactions with the interstellar medium (ISM).
Diffuse gamma rays are measured with high precision by \textit{Fermi}-LAT in the \si{\giga\electronvolt} range~\cite{2012AckermannAjelloApJ} and recently up to hundreds of \si{\tera\electronvolt} by LHAASO~\cite{2023CaoAharonianPhRvL, 2025CaoAharonianPhRvL}.
IceCube has also measured GDEs of neutrinos~\cite{2023IcecubeCollaborationAbbasiSci}.
The diffuse gamma rays are produced by CRs of roughly one order of magnitude higher energies~\cite{2006KelnerAharonianPhRvD,2021KoldobskiyKachelriessPhRvD}, thus providing insight into the high-energy end of Galactic CRs.
GDE intensities are derived by solving the CR transport equation with adequate transport parameters and source models to predict the CR intensities across the Galaxy, modelling the ISM properties, and using relevant cross sections.
The uncertainties associated with most of these ingredients are discussed in~\cite{2023SchweferMertschApJ}.
In this contribution, we focus on the influence of source models on GDE predictions, extending previous work by incorporating the discrete nature of sources.
A more detailed discussion of this work can be found in our recent publication~\cite{2025StallMertscharXiv}.

Supernova remnants (SNRs) are widely regarded the blueprint for sources of CRs in our Galaxy.
They accelerate particles via diffusive shock acceleration~\cite{1977KrymskiiSPhD}.
While it is debatable whether CRs can be accelerated up to the CR knee by SNRs, it is a common assumption that they account for most of the Galactic CRs~\cite{2019GabiciEvoliIJMPD}.
This is sometimes referred to as the \textit{supernova paradigm}.
With relatively small spatial and temporal scales compared to CR propagation, SNRs inject CRs basically as point sources.
However, the exact positions, ages, and properties of the CR sources that have contributed to the CR intensity today are mostly unknown.
Thus, a model to study the influence of the discrete nature of CR sources must be stochastic~\cite{2011MertschJCAP,2017GenoliniSalatiA&A, 2021EvoliAmatoPhRvDa} by considering the ensemble of possible source realisations.
We refer to this approach as \textit{stochastic modelling of sources}.
As an alternative, smooth and steady distributions are frequently used, but previous studies~\cite{2015KachelriessNeronovPhRvL,2025StallLooApJL} show that explicitly resolving discrete sources can modify CR intensity predictions, particularly at high energies where fewer sources contribute to the total CR intensity.

\begin{figure}
    \centering
    \includegraphics[width=\textwidth]{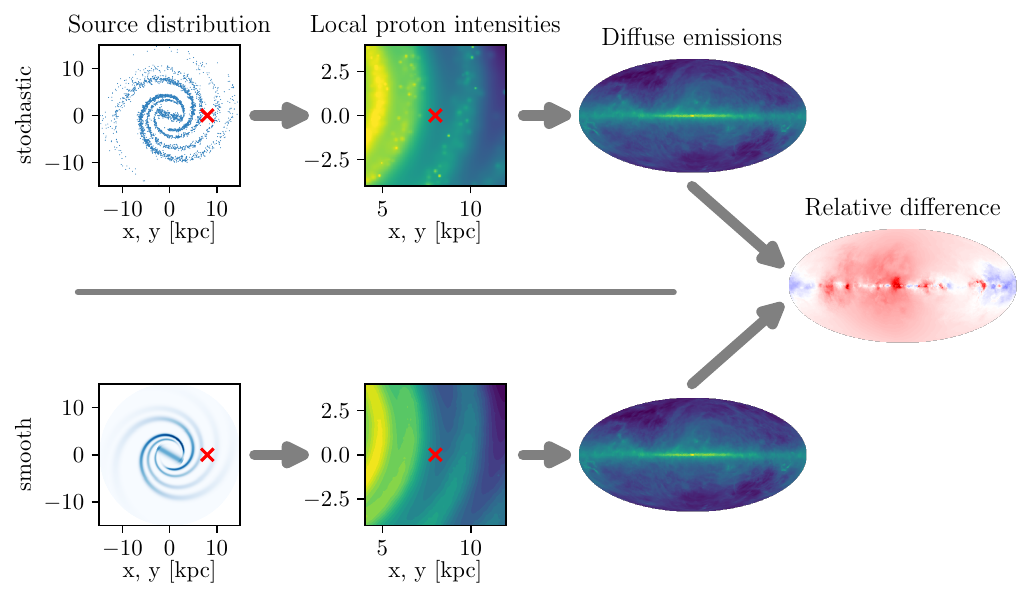}
    \caption{Schematic overview of stochastic vs. smooth source models.
    (This plot is also used in~\cite{2025StallMertscharXiv}.)}
    \label{fig:overview}
\end{figure}

In this contribution, we study the differences of GDE predictions in the case of stochastic source modelling and the predictions of the commonly assumed smooth and steady source distribution.
We call the effects of the discrete sources on the CR intensity and the GDEs \textit{source stochasticity}.
As local CR observations are not directly linked to CR intensities in other parts of the Galaxy, GDEs cannot be predicted based on local measurements of the CR intensity alone.
Inversely, measurements of GDEs can provide valuable insights on the distribution of CR sources and their properties in other parts of the Galaxy.
It has been shown~\cite{2017GenoliniSalatiA&A} that the CR intensity at any given position in the Galaxy approximately follows a stable distribution which exhibit power-law tails.
We build on this by investigating how discrete sources affect the distribution of GDEs.
Several authors have discussed discrete source effects on gamma-ray and neutrino intensities~\cite{2023ThalerKissmannAPh,2023MarinosRowellMNRAS,2025MarinosPorterApJ,2025KaciGiacintiJCAP} with small sample sizes.
We want to capture the effects of source stochasticity and source properties with extensive Monte Carlo simulations.
Fig.~\ref{fig:overview} illustrates our workflow and contrasts it to the one in smooth source modelling:
We draw discrete sources from a smooth distribution, compute the resulting CR proton intensities throughout the Galaxy, and derive GDE sky maps.
By comparing these with smooth source model predictions, we quantify the effect of discrete sources on the GDE predictions.
\section{Methods}
We present our stochastic model for calculating hadronic GDEs by modelling the injection of CR protons from discrete sources.
Numerically efficient methods are crucial for analysing many source realisations in a Monte Carlo framework.
Hence, we also focus on efficient computational methods for the calculation of CR proton intensities and the line-of-sight integration to predict GDEs.
\subsection{Transport model}\label{sec: Transport model}
We assume that CR sources are distributed according to a spiral model (see~\cite{2021EvoliAmatoPhRvDb,2021EvoliAmatoPhRvDa,2010Steiman-CameronWolfireApJ} adopting a solar radius $R_{\odot} = \SI{8}{\kilo\parsec}$ and the radial distribution from~\cite{2001FerriereRvMP}) near the Galactic disk ($z=0$) with a Gaussian distribution with a scale height of $z_0 = \SI{70}{\parsec}$ in the $z$-direction.
The source rate is set to \SI{0.03}{\per\year}~\cite{1994TammannLoefflerApJS}.
We also assume that CR protons diffuse in the Galactic halo that has free-escape boundaries at $z=\pm H$.
Instead of the CR energy, we use rigidity, defined as $\mathcal{R} = {p c}/{Z e}$ where $p$ denotes the particle's momentum, $c$ the speed of light, $Z$ the charge number, and $e$ the unit charge.
For CR protons with rigidities above \SI{10}{\giga\electronvolt}, the transport equation for the isotropic CR density $\psi_\mathcal{R}(\mathcal{R}, t, \mathbf{x})$ ($ = \mathrm{d}n/\mathrm{d}\mathcal{R}$, where $n$ is the number density) simplifies to a diffusion equation:
\begin{equation}\label{eq:transport_equation}
    \frac{\partial \psi_{\mathcal{R}}\left(\mathcal{R}, t, \mathbf{x}\right)}{\partial t} \;-\; \kappa\left(\mathcal{R}\right) \nabla^2 \psi_{\mathcal{R}} \left(\mathcal{R}, t, \mathbf{x}\right) \;=\; Q\left(\mathcal{R}, t, \mathbf{x}\right) \, ,
\end{equation}
where $\kappa(\mathcal{R})$ is the (isotropic) diffusion coefficient, and $Q(\mathcal{R},t,\mathbf{x})$ is the source term describing the CR injection.
Both follow the form described in~\cite{2023SchweferMertschApJ}.
The prediction for the total CR proton intensity is obtained by summing the solutions of this equation (incl. boundary conditions), i.e., the Green's functions, for millions of sources.
For more details, see~\cite{2025StallLooApJL}.
\subsection{Source injection and near-source transport}\label{sec: source injection and near-source transport}
We allow for (i) simultaneous (burst-like) injection or (ii) a rigidity-dependent escape time (CREDIT~\cite{2025StallLooApJL}), as well as (iii) time-dependent diffusion (TDD) to capture a phase of reduced diffusion around young sources~\cite{2025KaciGiacintiJCAP}.
All scenarios are implemented by adjusting the injection time and diffusion coefficient in the respective Green's functions.
\subsection{Diffuse emissions and absorption}\label{sec: diffuse emissions}
The intensity of hadronic GDEs at energy $E$ in direction $(l,b)$ is given by a line-of-sight integral of the CR proton intensity $\Phi(\mathbf{x}, E')$ multiplied by the appropriate cross sections and gas densities:
\begin{equation}\label{eq: diffuse emissions general}
    J(l, b, E) = \frac{1}{4 \pi} \int_0^{\infty} \mathrm{d} s \ e^{- \tau\left(E, \boldsymbol{x}\right)} \ n_{\text{gas}}(\boldsymbol{x}) \int_E^{\infty} \mathrm{d} E' \ \frac{\mathrm{d} \sigma}{\mathrm{d} E}(E', E) \ \Phi(\boldsymbol{x}, E') \Big|_{\boldsymbol{x} = \boldsymbol{x}(l, b, s)}
\end{equation}
where $n_{\text{gas}}$ is inferred from HI and CO surveys~\cite{2025SodingEdenhoferA&A}, \textsc{AAfrag} cross sections~\cite{2019KachelriessMoskalenkoCoPhC,2021KoldobskiyKachelriessPhRvD} are used, and scaling factors account for heavier CRs and gas components~\cite{2015CasandjianApJ,2004HondaKajitaPhRvD}.

Above \si{\tera\electronvolt} energies, gamma-ray absorption via pair absorption $\gamma\gamma\rightarrow e^+e^-$ can be significant, mainly due to infrared target radiation fields from dust~\cite{2018PorterRowellPhRvD} and the CMB.
We incorporate precomputed absorption coefficients from \textsc{GALPROP}~\cite{2022PorterJohannessonApJS}, interpolated onto our spatial and energy grid.
The resulting optical depth $\tau(E, \mathbf{x})$ enters Eq.~\eqref{eq: diffuse emissions general}, attenuating high-energy gamma rays.
\subsection{Computational strategy}
The CR intensity summation over millions of sources and the line-of-sight integrals are parallelised using graphical processing units (GPUs).
We do this by using the \textsc{python} package \textsc{jax}~\cite{jax2018github}.
This gives us a significant speed-up compared to (parallelised) \textsc{python} implementations.
We also split young and old sources, with old sources contributing a fixed background, further reducing the computational demand.
This approach allows us to study large numbers of source realisations, increasing the statistical strength of our results.
\section{Results}\label{sec:results}
\begin{figure}
    \centering
    \includegraphics[width=\textwidth]{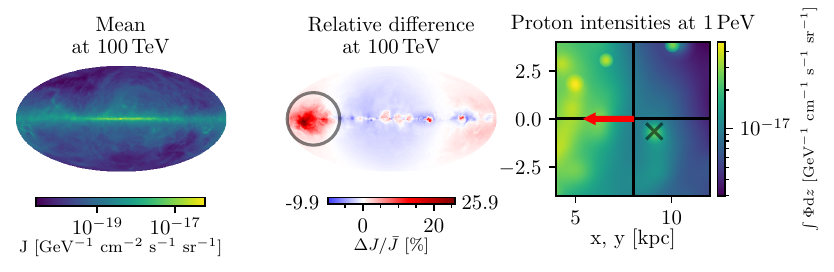}
    \caption{These figures illustrate the influence of young sources on GDE predictions.
    The mean predictions for the GDEs at \SI{100}{\tera\electronvolt} are shown on the left.
    Between the relative difference of one specific realisation from this mean (centre) and the integrated local CR proton intensities (right), correspondences can be found (e.g., as indicated).
    The red arrow points towards the Galactic centre which is in the centre of the sky maps.}
    \label{fig:local}
\end{figure}
\begin{figure}
     \centering
     \begin{subfigure}[b]{0.49\textwidth}
         \centering
         \includegraphics[width=\textwidth]{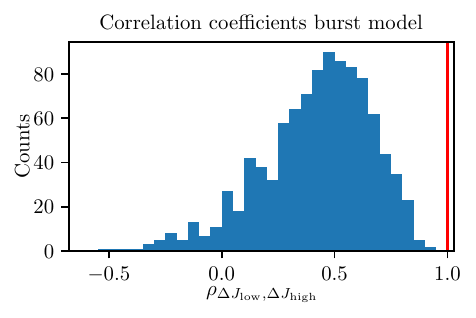}
     \end{subfigure}
     \hfill
     \begin{subfigure}[b]{0.49\textwidth}
         \centering
         \includegraphics[width=\textwidth]{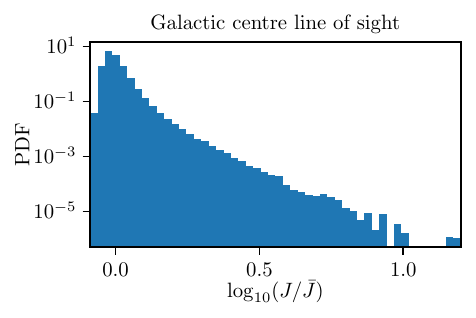}
     \end{subfigure}
        \caption{Results of the statistical analysis for the burst-like injection model are shown.
        Histogram of Pearson correlation coefficients (left) and histogram of GDE intensity deviations along the Galactic centre line of sight for a sample size of $10^6$ realisations (right).}
        \label{fig3: results}
\end{figure}

We have simulated $1\,000$ realisations of GDEs of the full sky at $24$ different GDE energies for each of the three source injection and near-source transport models described in section~\ref{sec: source injection and near-source transport}.
We find a clear correspondence between higher-than-average CR intensities from young sources and over-fluctuations in the GDEs.
Fig.~\ref{fig:local} illustrates this effect for a GDE energy of \SI{100}{\tera\electronvolt}, where regions with significant local CR enhancements align with stronger GDE intensities.
By comparing relative difference sky maps at \SI{10}{\giga\electronvolt} (hereafter \textit{low} energy) and at \SI{100}{\tera\electronvolt} (\textit{high}), we find that the correspondence between regions of higher or lower deviations from the mean depends strongly on the source injection and near-source transport models described in section~\ref{sec: source injection and near-source transport}.
In particular, we observe that the correlation between low- and high-energy GDEs can be partially or entirely broken (see Fig.~\ref{fig3: results}~(left)).
While a morphological correspondence between the sky maps at low and high energies is present in the burst as well as in the TDD scenario, we find that it would be misleading to assume that GDEs at low energies can be extrapolated to high energies.

The highest relative differences in GDE intensities across the sky can reach tens of percent in the burst and CREDIT model.
For the TDD scenario, extreme deviations of hundreds or even thousands of percent occur due to the long confinement of CRs close to their sources and the resulting strong enhancement of CR intensities around these sources.

For selected lines of sight, we have increased our statistical power and simulated $10^6$ realisations for each scenario and GDE energy.
Thanks to this large statistical sample, we can study the distributions of the fluctuations of realisations around the mean.
In Fig.~\ref{fig3: results}~(right), we can see a resulting histogram of the line of sight towards the Galactic centre that can be compared to the ones for the local CR proton intensity shown in~\cite{2017GenoliniSalatiA&A}.
The distribution shows a clear power-law tail.
\section{Conclusion}
We end the short presentation of the results of the analysis of our Monte Carlo simulations with the following conclusions.
More details and discussion can be found in Ref.~\cite{2025StallMertscharXiv}.
\begin{enumerate}
    \item Considering discrete CR sources leads to considerable deviations from the GDEs expected from a smooth source distribution.
    The strength of these deviations can depend a lot on the source injection and near-source transport model.
    \item The diffuse sky at low and high energies is in general not well correlated.
    \item Monte Carlo simulations sped up by GPUs make it possible to study GDE distributions with large sample sizes.
\end{enumerate}
\acknowledgments
This work has been funded by the Deutsche Forschungsgemeinschaft (DFG, German Research Foundation) — project number 490751943.
The authors also gratefully acknowledge the computing time provided to them at the NHR Center NHR4CES at RWTH Aachen University (project number p0021785).

\bibliographystyle{JHEP}
\bibliography{references.bib}

\end{document}